\documentclass[useAMS,usenatbib]{mnras}
\usepackage{graphicx}
\usepackage{color}
\usepackage{tikz}
\usepackage{breakurl}
\usepackage{setspace}
\usepackage{amssymb}
\usepackage{comment}
\usepackage{xcolor}
\usetikzlibrary{positioning}
\usetikzlibrary{matrix,arrows}
\title[SCORPIO-II: Spectral indices of weak Galactic radio sources]
  {SCORPIO-II: Spectral indices of weak Galactic radio sources}
\author[F. Cavallaro et al.] {F.~Cavallaro$^{1,2,3}$, C.~Trigilio$^2$, G.~Umana$^2$, T.M.O.~Franzen$^4$,  R.P.~Norris,$^{3,5}$,  P.~Leto$^2$, \newauthor A.~Ingallinera$^{2}$,  C. S.~Buemi$^2$, J. Marvil$^3$, C.~Agliozzo$^{6,7}$, F. Bufano$^2$, L.~Cerrigone$^8$, \newauthor 
S. Riggi$^2$ \\\\
  $^1$Universit\`a degli studi di Catania, Catania, Italy\\
  $^2$INAF- Osservatorio Astrofisico di Catania, Via S. Sofia 78,  95123, Catania, Italy\\
  $^3$CSIRO Astronomy and Space Science, PO Box 76, Epping, NSW 1710, Australia\\
  $^4$International Centre for Radio Astronomy Research, Curtin University, Bentley, WA 6102, Australia\\
  $^5$Western Sydney University, Locked Bag 1797, Penrith South, NSW 1797, Australia\\
  $^6$Millennium Institute of Astrophysics, Santiago, Chile\\
  $^7$Universidad Andr\'es Bello, Avda. Republica 252, Santiago, Chile\\
  $^8$ASTRON, the Netherlands Institute for Radioastronomy, PO Box 2, 7990 AA Dwingeloo,The Netherlands\\
  }
 
\begin{document}
\date{Draft Oct 2016}

\maketitle
\label{firstpage}

\begin{abstract}
In the next few years the classification of radio sources observed by the large surveys will be a challenging problem, and spectral index is a powerful tool for addressing it. Here we present an algorithm to estimate the spectral index of sources from multiwavelength radio images. We have applied our algorithm to SCORPIO \citep{SCORPIO}, a Galactic Plane survey centred around 2.1 GHz carried out with ATCA, and found we can measure reliable spectral indices only for sources stronger than 40 times the rms noise. Above a threshold of 1 mJy, the source density in SCORPIO is 20 percent greater than in a typical extra-galactic field, like ATLAS \citep{ATLAS}, because of the presence of Galactic sources. Among this excess population, 16 sources per square degree have a spectral index of about zero, suggesting optically thin thermal emission such as H\textsc{ii} regions and planetary nebulae, while 12 per square degree present a rising spectrum, suggesting optically thick thermal emission such as stars and UCH\textsc{ii} regions. 
\end{abstract}


\section{Introduction}

We are entering a golden age for radio astronomy. There are several new interferometers upcoming, including the Square Kilometre Array (SKA; \citealt{carilli}) and its precursors, such as the Australian SKA Pathfinder (ASKAP, \citealt{johnston}) and MeerKAT \citep{jonas}, that will deliver an extraordinary number of new discoveries. In particular, we expect a lot of data from the forthcoming radio continuum surveys, both in the Galactic Plane (GP) and at high Galactic latitude, such as the Evolutionary Map of the Universe (EMU; \citealt{EMU}), to be carried out with ASKAP, and the MeerKAT International GigaHertz Tiered extra-galactic Exploration Survey (MIGHTEE; \citealt{jarvis}) and the MeerKAT High Frequency Galactic Plane Survey (MeerGAL), both to be carried out with MeerKAT. We will need to classify a large number of sources, especially in the GP, where the Galactic population adds to the extragalactic population. There are essentially six ways to classify a generic radio source, depending on the data available and whether it is resolved or not: 
\begin{enumerate}
	\item studying its morphology, which requires that the source is resolved;
	\item studying the radio emission mechanism, which requires that the spectral energy distribution (SED) in the radio wavelengths is known;
	\item studying the complete SED, which requires the identification of the source counterparts at different wavelengths;
	\item studying the polarization, which requires a very high signal-to-noise ratio (S/N);
	\item studying the time domain, which requires a sufficient time resolution;
\end{enumerate} 
In this paper we will focus on the second method, since it is solely based on radio multiwavelength or wide-band observations and no other information is required. The radio SED depends on the continuum emission mechanism, which could mainly be thermal (bremsstrahlung) or non-thermal (synchrotron).

A parameter that characterises the SED at the radio wavelengths is the spectral index $\alpha$, defined as $S=S_0\left(\frac{\nu}{\nu_0}\right)^\alpha$, where $S$ is the flux density of the source at a given frequency $\nu$ and $S_0$ is the flux density at the frequency $\nu_0$. The spectral index can change with the frequency, but both synchrotron and bremsstrahlung lose this dependence in the two optical depth limits $\tau\gg1$ and $\tau\ll1$. In particular, in the case of a homogeneous source of bremsstrahlung we have $\alpha=2$ for $\tau\gg1$ while, for $\tau\ll1$, $\alpha=-0.1$ \citep{burke234}. In the case of $n\propto r^{-2}$ (like a stellar wind), where $n$ is the electronic density and $r$ is the distance from the centre, $\alpha=0.6$ (\citealt{Wright}; \citealt{panagia}). In the case of a homogeneous source of synchrotron $\alpha=2.5$ for $\tau\gg1$ while, for $\tau\ll1$, $\alpha$ depends on the spectral index $\delta$ of the energy distribution of the relativistic electrons, defined as $\alpha=-\frac{\delta-1}{2}$ so, for typical values of $\delta$, it ranges between $-1$ and $-0.5$ \citep{burke131}. 

The SED can also be modified by synchrotron self-absorption and free-free absorption, causing a low frequency turnover, and by electron cooling, causing a high frequency break. Complex models are invoked to reconcile these models with the data, which are beyond the scope of this paper, but they are discussed extensively by Collier et al., in preparation.

The frequency at which the flux density of a source reaches a maximum is called the turnover frequency. Measuring the turnover frequency can allow us to constrain additional physical parameters, such as magnetic field intensity or electronic density. To derive the spectral index and all the connected informations for a large number of sources, considering that the future surveys will detect millions of sources \citep{EMU}, we developed an algorithm to automatically compute the spectral index. 

Our algorithm has been tested by using the images of Stellar Continuum Originating from Radio Physics In Ourgalaxy (SCORPIO, \citealt{SCORPIO}, hereafter Paper I) survey. SCORPIO is a $2\times2\,\mathrm{deg}^2$ survey in the GP centred at Galactic coordinates $l=344°.25$, $b=0°.66$, carried out with the Australian Telescope Compact Array (ATCA) between 1.1 and 3.1 GHz that acts as a pathfinder for EMU. A pilot experiment took place in 2011 and in 2012, observing about a quarter of the selected field. The rms of the map was $\approx25-30\,\mathrm{\mu Jy/beam}$, with a resolution of 14.0 by 6.5 arcsec (Paper I). Notice that SCORPIO is actually larger than 4 deg$^2$ because, to have a better coverage of the region, we preferred to observe a larger area. This can also be applied to the pilot, in fact its area is actually $2.333$ deg$^2$, compared to the declared 1 deg$^2$.

In Paper I we produced a catalogue of 614 point sources in the pilot field, listing their flux densities, position and possible infrared counterpart. The data release of the full SCORPIO field, with a catalogue of about 2000 sources, is in progress (Trigilio et al., in preparation). In this paper we present a statistical analysis of the radio SED of the 614 point sources extracted and catalogued in Paper I, to statistically discriminate Galactic from extra-galactic point sources. In Section 2 we give a brief overview of the observations and data reduction. In Section 3 we describe our algorithm to automatically reconstruct the source SED and derive the spectral index. In Section 4 we present our results, thoroughly discussing their relation with ATLAS (Australia Telescope Large Area Survey, \citealt{ATLAS}) ones, highlighting and explaining the differences. Conclusions are reported in Section 5. 
\section{Data}
The observations were performed in the periods 21-24 April 2011 and 3-11 June 2012, using ATCA in 6A and 6B configurations at 16-cm. The field was observed using a mosaic of 38 pointings. A detailed description of the observations and data reduction is reported in Paper I. Here we briefly describe the main procedures we followed to produce the final images. 
\subsection{Data cube generation}
Flagging was performed in \texttt{MIRIAD} \citep{Miriad}, using the task \texttt{mirflag}. Radio Frequency Interference (RFI) was removed by flagging 30 to 40 per cent of the visibilities. The usable range of frequencies covers 1.350 to 3.100 GHz, corresponding to $\Delta\nu/\nu\approx0.8$. Calibration was performed using the standard \texttt{MIRIAD} tasks and the standard bandpass calibrator, the radio galaxy 1934-638, which had an assumed flux density of 12.31 Jy at 2.1 GHz \citet{Reynolds}. We divided our data in 7 sub-bands (see Table \ref{sub}).

The imaging was performed in \texttt{MIRIAD}. The deconvolution was executed with the \texttt{mfclean} task using the H\"{o}gbom algorithm \citep{Hogbom} on every pointing. The cell size was set to 1.5 arcsec and the restored Gaussian beam was set to 14.0 by 6.5 arcsec. The pointings were then joined into one mosaic using the \texttt{linmos} task. Finally we obtained 7 maps at 7 different contiguous frequency bands. This division is very convenient to calculate the SED of the sources, considering that the 16-cm band is often the turn-over region for galactic sources. An all-band map has been obtained using all the uv data to accomplish the best S/N.

The source extraction was conducted on a region of the all-band map as in \citet{Franzen}. We chose to restrict our studies to the largest region with a uniform and low rms, by taking into account the size of the primary beam at the highest frequency sub-band, given that at this frequency the primary beam is the smallest (see paper I). The maps of the local noise were used to identify sources on the basis of their S/N. Local maxima above $5\sigma$ were identified as sources. A peak position and flux density value were measured by interpolating between pixels. A centroid position, integrated flux density and source area were also calculated by integrating contiguous pixels down to $2.5\sigma$ (as in \citealt{waldram}), and sources were identified as overlapping if the integration area contained more than one source. Our source extraction process yielded a sample of 614 point sources. 


While the majority of the sources extracted by the \citet{Franzen} algorithm are point-like, a few of them are slightly resolved. The most extended one is SCORPIO\_169a, with an area of 8.3 times the beam area (Paper I). This source has a linear angular scale of $\sim35''$, while our Largest Angular Scale is $\sim2.3'$, so we are sensitive to scales significantly larger than our largest compact source.

The study of SCORPIO extended sources is beyond the scope of this paper, but we are using the extraction algorithm presented in \citet{Riggi} and analysing them in Ingallinera et al., in preparation.
\begin{table}
\begin{center} 
\begin{tabular}{l | l | l}
\hline
Sub-bands & Central frequency  & $R$ \\
&(GHz)&arcmin\\
\hline
1&1.469&30.8\\
2&1.649&27.5\\
3&1.850&24.4\\
4&2.075&23.6\\
5&2.329&20.1\\
6&2.613&18.7\\
7&2.932&16.7\\
\hline
\end{tabular}
\caption{Sub-band frequency and primary beam radii at 5 percent of the peak.}\label{sub}
\end{center}
\end{table}
\subsection{Primary beam correction}
The mosaicing process requires the correction of the telescope primary beam in order to get accurate brightness in the field. A non accurate knowledge of the primary beam prevents a good flux density measurement, in particular for the sources far from the center of the individual pointings. As a consequence, the spectral indices of the sources will be affected non-uniformly in the field.

In the last few years measurements of the primary beam shape for the new 16-cm CABB receivers were made across the entire usable frequency range (1.1 to 3.1 GHz). Although an accurate primary beam model is now available, it had not been implemented in MIRIAD when SCORPIO imaging was performed. 

Instead, in Paper I, we used the Gaussian primary beam model made by \citet{Wieringa} (see Table \ref{corr}) (Paper I). The product of the FWHM and the frequency $d^\prime$, is, according to them, constant for the first three sub-bands, then it changes and remains constant for the last 4 ones because they used different receivers, where the primary beam is assumed to vary as $\nu^{-1}$ in a single receiver. New primary beam measurements show that this is not true\footnote{\url{http://www.narrabri.atnf.csiro.au/people/ste616/beamshapes/beamshape\_16cm.html}} (in Table \ref{corr}, the new ones are reported as $d$ values).  The primary beam FWHM tends to decrease less rapidly then $\nu^{-1}$ below 2.1 GHz. It can be caused by a deterioration in the focus above 1.4 GHz. 

In this paper, we use the new version of MIRIAD, in which \texttt{LINMOS} has the right primary beam correction, so we decided to redo the mosaicking, applying the new \texttt{LINMOS} task to the pointings to obtain the corrected maps. The slight differences in the fluxes between the data in this paper and the data in Paper I is due to this correction.
\begin{table}
\begin{center} 
\begin{tabular}{l | l | l}
\hline
$\nu$ & $d$ & $d^\prime$ \\
(GHz)& (arcmin GHz) & (arcmin GHz)\\ 
\hline
1.469 & 48.34 &47.90\\
1.649 & 49.34 &47.90\\
1.850 & 50.40 &47.90\\
2.075 & 51.03 &49.70\\
2.329 & 49.91 &49.70\\
2.613 & 49.54 &49.70\\
2.932 & 48.31 &49.73\\
\hline
\end{tabular}
\caption{Product of the primary beam FWHM and frequency at every SCORPIO sub-band, using the more recent measurements of the primary beam for the new 16-cm CABB receiver ($d$) and the primary beam model by Wierenga \& Kesteven (1992) ($d^\prime$).}\label{corr}
\end{center}
\end{table}

\section{Image analysis}\label{alg}
Each of our 614 sources is characterised by a SED that provides us information on the mechanism of the radio emission (thermal or non thermal), and, if we find the turn-over frequency, some physical parameter such as magnetic field or electron density. This information can help us to discriminate between galactic and extra-galactic sources and to characterise the source, giving us hints on its nature. 
\subsection{Flux Density Extraction Algorithm} \label{fluxextractor}
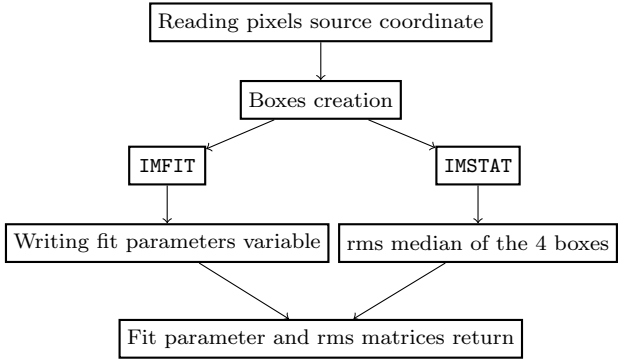
\begin{figure}
\begin{center}
\begin{tikzpicture}[mystyle/.style={draw,rectangle,fill=white!30,thick,minimum
width=1cm,minimum height=0.5cm}]
\node[mystyle] (A) {Reading pixels source coordinate};
\node[mystyle,node distance=0.5cm] (B) [below=of A] {Boxes creation};
\node[minimum width =1cm,node distance=0.5cm] (D) [below=of B] {};
\node[minimum width =1cm,node distance=0.5cm] (I) [below=of D] {};
\node[minimum width =1cm,node distance=0.5cm] (L) [below=of I] {};
\node[mystyle,node distance=0.5cm] (M) [below=of L] {Fit parameter and rms matrices return};
\node[mystyle,node distance=1cm] (E) [left=of D] {\texttt{IMFIT}};
\node[mystyle,node distance=1cm] (F) [right=of D] {\texttt{IMSTAT}}; 
\node[mystyle,node distance=0.5cm] (G) [below=of E] {Writing fit parameters variable}; 
\node[mystyle,node distance=0.5cm] (H) [below=of F] {rms median of the 4 boxes};

\draw[->] (A) -- (B);
\draw[->] (B) -- (E);
\draw[->] (B) -- (F);
\draw[->] (E) -- (G);
\draw[->] (F) -- (H);
\draw[->] (G) -- (M);
\draw[->] (H) -- (M);

\end{tikzpicture}
\end{center}
\caption{Flowchart for the flux density extraction algorithm}\label{schemagauss}
\end{figure}

In order to automatically estimate the source flux densities at different frequencies, we wrote a Python script that runs in CASA (Common Astronomy Software Applications, \citealt{CASA}) and uses the \texttt{imfit} task (see flowchart in Fig. \ref{schemagauss}). The script reads the position of each source from the Paper I catalogue and creates a box around it and 4 other boxes, that are 15 pixels along the x axis and 10 pixels along the y axis away from the vertices. The box dimensions are chosen to be 3 times the beam major and minor axis respectively (the major axis is oriented along the north-south direction). The script then calls \texttt{imfit} in the central box and \texttt{imstat} in the other four ones. \texttt{imfit} performs a bidimensional Gaussian fit in the chosen region and returns a Python dictionary $dic$ containing all the fit parameters (such as integrated flux density, Gaussian peak position, major and minor axis dimension, etc.). The \texttt{imfit} task implements an algorithm that models and subtracts the background from the fit thereby removing the effect of diffuse emission. The script uses \texttt{imstat} to measure the rms and median noise in	each of the adjacent boxes, and then calculates the median background noise. At the end the script returns two matrices with $N$ lines and $M$ columns, where $N$ is the number of sources and  $M$ the number of sub-bands. Every element $a_{ij}$ of the first matrix is the fit parameter $dic$ of the $i^{\mathrm{th}}$ source for the $j^{\mathrm{th}}$ sub-band, in a similar way every element $b_{ij}$ of the second one is the local rms of the $i^{\mathrm{th}}$ source for the $j^{\mathrm{th}}$ sub-band. If the Gaussian fit does not converge for the $i^{\mathrm{th}}$ source and the $j^{\mathrm{th}}$ sub-band, $a_{ij}$ is simply a $0$ (see Fig. \ref{schemagauss}).

To check the reliability of the algorithm, we applied it to the old all-band map and compared the results with the source flux densities extracted using the \citet{Franzen} algorithm. The value of the average ratio $r=S_\mathrm{f}/S_{c}$, where $S_\mathrm{f}$ is the flux density measured by the \citet{Franzen} algorithm and $S_\mathrm{c}$ is the one measured using our algorithm, is $r=0.97\pm0.13$. This value shows that the two methods are consistent.

We also ran our algorithm on the new all-band map, to check for differences with the old map flux densities. We found a ratio of $r=0.94\pm0.44$ between the flux densities measured with our algorithm in the new all-band map and $S_\mathrm{c}$. This higher difference is consistent with the primary beam correction.
\subsection{Spectra construction and control algorithm}
For the purposes of this work, we initially assume that all sources have a power-law spectrum. Our algorithm performs a weighted linear fit of $\log{S}$ as a function of $\log\nu$, where $S$ is the flux density and $\nu$ the frequency and the single point error is \begin{eqnarray}\Delta S=\sqrt{e_\mathrm{fit}^2+e_\mathrm{rms}^2+e_\mathrm{cal}^2}\end{eqnarray} where $e_\mathrm{fit}$ is the fit uncertainty, computed by \texttt{imfit}, $e_\mathrm{rms}$ is the noise and $e_\mathrm{cal}$ is the calibration error. To check that the SED has a linear behaviour it fits a second degree polynomial and compares the chi squared ($\chi^2$) of the linear and parabolic fits. In the 30 sources in which the $\chi^2$ of the parabolic fit has a lower value, we ignored the low-frequency turnover by excluding all data at a frequency below that of the peak and redo the fit to get $\alpha$. 

\begin{figure}
\begin{center}
\begin{tikzpicture}[mystyle/.style={draw,rectangle,fill=white!30,thick,minimum
width=3cm,minimum height=0.5cm},mystyle2/.style={draw,rectangle,fill=white!30,thick,minimum width=3cm,minimum height=0.5cm},mystyle3/.style={draw,circle,fill=red!250,thick}]
\node[mystyle2] (A) {Input: Sources fluxes and errors};
\node[mystyle2,node distance=0.8cm] (B) [below=of A] {Checking for ``bad points''}; 
\node[mystyle2,node distance=0.8cm] (M) [below=of B]{Weighted Linear fit};
\node[mystyle2,node distance=0.8cm] (N) [below=of M]{Writing linear fit parameters};

\draw[->] (A) -- (B);
\draw[->] (B) -- (M);
\draw[->] (M) -- (N);

\end{tikzpicture}
\end{center}
\caption{Spectral index extraction algorithm flowchart.}\label{schemafit}
\end{figure}
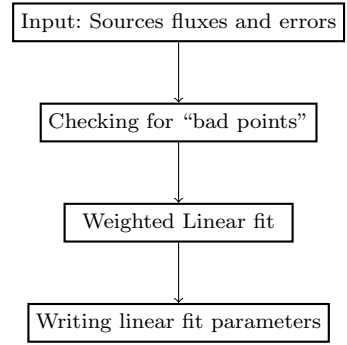

The algorithm performs the following tests for each source in each of the 7 sub-bands to flag ``bad points'' (see Fig. \ref{schemafit} for a simplified flow-chart of the procedure):
\begin{enumerate}
\item The flux density $S$ is equal or less than 0. This is the case of non-converged Gaussian fit or Gaussian fit of negative values on the map, for e.g. caused by artifacts: 
\begin{eqnarray}
S\le0,
\end{eqnarray}
$\sim36$ percent of the 4298 (614 sources multiplied by the 7 sub-bands) flux density measurements were flagged by this criterion;
\item the difference between the position of the fitting Gaussian peak and the position given by the source extraction algorithm is greater than half the width of the beam $w$. This is the case of a Gaussian fit made on a near brighter source with high sidelobes:
\begin{eqnarray}P_{\mathrm{peak}}-P_{\mathrm{list}}\ge \frac{w}{2}, \end{eqnarray}
$\sim6$ percent of the points were flagged by this criterion
\item The difference between the area of the fitting Gaussian at a given frequency and the average of the area of the other frequencies of the same source is greater than 2 times their standard deviation. This is the case of fitting a faint source that can be affected by noise: 
\begin{eqnarray}\left|A-\bar{A}\right|\ge 2\sigma,\end{eqnarray}
where $A$ and $\bar{A}$ are the area of the fitting Gaussian for a sub-band and the mean area of the other sub-bands respectively and $\sigma$ is its standard deviation. 

About 1 percent of the points were flagged by this criterion.
\item The flux density is significantly greater than the average $\bar{S}$. 
\begin{eqnarray}S>n\bar{S}. \end{eqnarray}
We tested several values of $n$ between 1 and 2 over $\sim50$ sources. The results have been visually checked and it turns out that $n=1.8$ was the minimum value above which the fit was unreliable.

About 9 percent of the points were flagged by this criterion.
\end{enumerate}
If a flux density measurement shows one of those irregularities, it is discarded and does not contribute to the linear fit.

\subsection{Grouped Components} \label{group}
A sub-sample of 65 of the 614 sources are classified by the source finding algorithm as ``double sources''\citep{Franzen}. They can be components of larger extended sources or simply be near to each other by projection. To fit these multiple components \texttt{imfit} needs to read a file that contains a list of estimates of the position of the Gaussians peaks, their dimensions, orientation and brightness. We built a script that automatically writes this file for each group of components, keeping the Gaussian dimensions and orientation fixed. Then the flux extractor algorithm works in the same way it works for the single sources. The algorithm treats each component of a double source as a real component of a larger source, adding the integrated flux densities and using the result as the flux density of the complex source.
\subsection{MGPS-2 matches}

\begin{figure*}
	\begin{center}
		\includegraphics[scale=0.65]{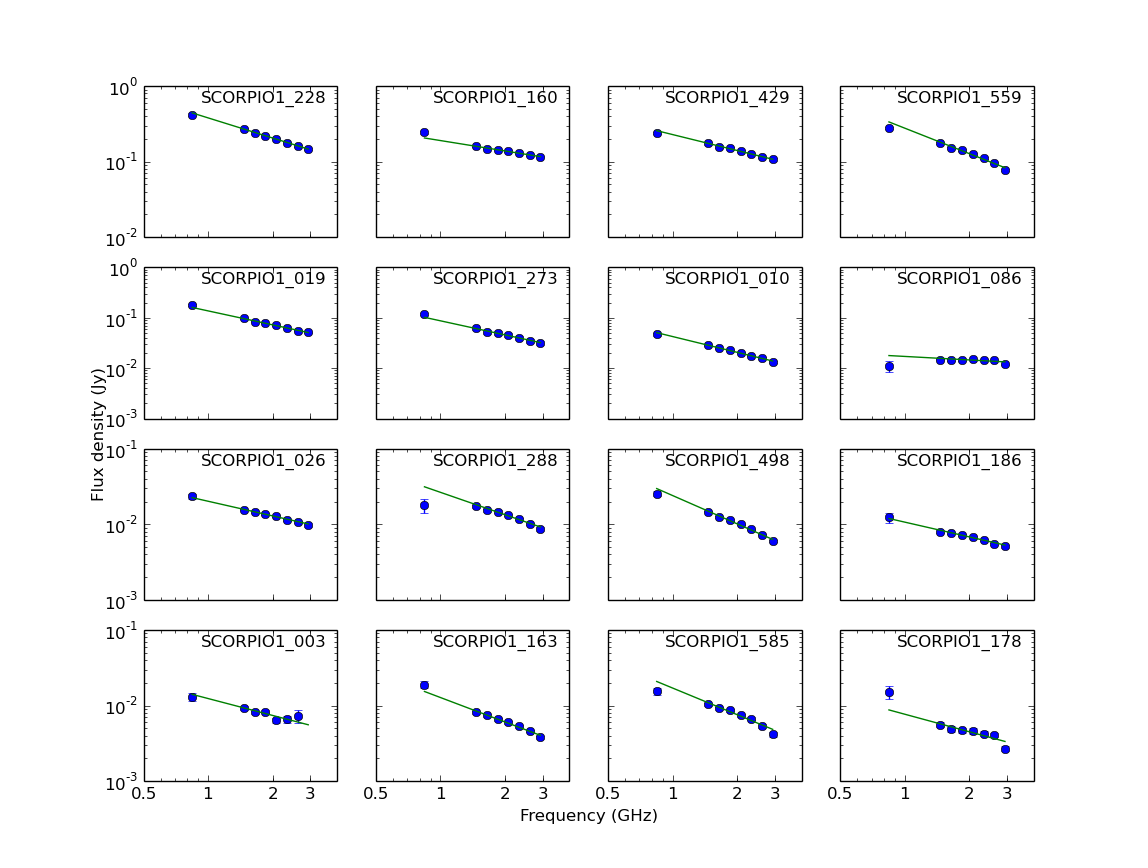}
	\end{center}
	\caption{Spectra of the 16 sources selected among the 43 sources that match with the MGPS-2.}	\label{MGPS2}
\end{figure*}

\begin{figure*}
	\begin{center}
		\includegraphics[scale=1]{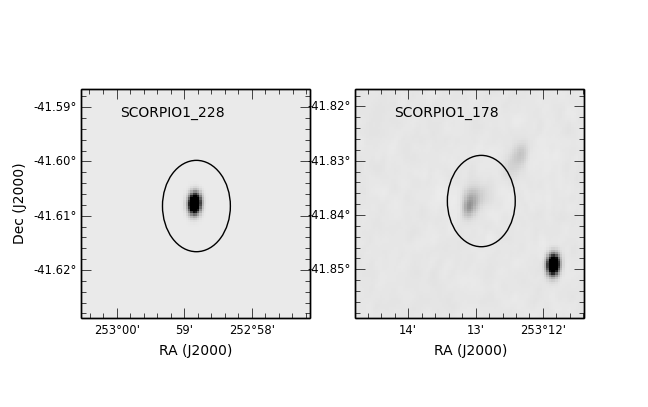}
	\end{center}
	\caption{Example of SCORPIO sources with a MGPS-2 match. The Molonglo beam is reported as the black ellipse to stress the effects of a large beam size in flux density measurements.}	\label{MGPS2im}
\end{figure*}

To check our spectral indices we considered the 43 SCORPIO sources that match with the second epoch Molonglo Galactic Plane Survey (MGPS-2, \citealt{MGPS}) sources (Paper I). MGPS-2 is a survey of the Galactic Plane, carried out with the Molonglo Observatory Synthesis Telescope (MOST) at a frequency of 843 MHz and with a restoring beam of $45'' \times 45'' \csc\delta$, where $\delta$ is the declination so it is suited to expand the SCORPIO sources' SED and check for spectral index errors. If we exclude the double sources and those embedded in diffuse emission (with confusion, because of the large MGPS-2 beam), there are only 16 sources left. We calculated their spectral index, including the 843-MHz MGPS-2 point. In Fig. \ref{MGPS2} we show the spectra of the 16 sources. Except for SCORPIO1\_178 and SCORPIO1\_288 the SEDs are consistent with the 843 MHz flux density. In the case of SCORPIO1\_178 the discrepancy is due to the fact that the MGPS-2 centroid is not centred on the SCORPIO source and its beam encompasses another brighter source, leading to a higher flux density, while in the case of SCORPIO1\_288 it can be due to a turn-over effect. In Fig. \ref{MGPS2im} we show the effect of the large beam size of MGPS-2 in the case of an isolated source (left panel) and SCORPIO1\_178 (right panel). 
\section{Spectral indices analysis} \label{data}
\begin{figure}
	\begin{center}
		\includegraphics[scale=0.45]{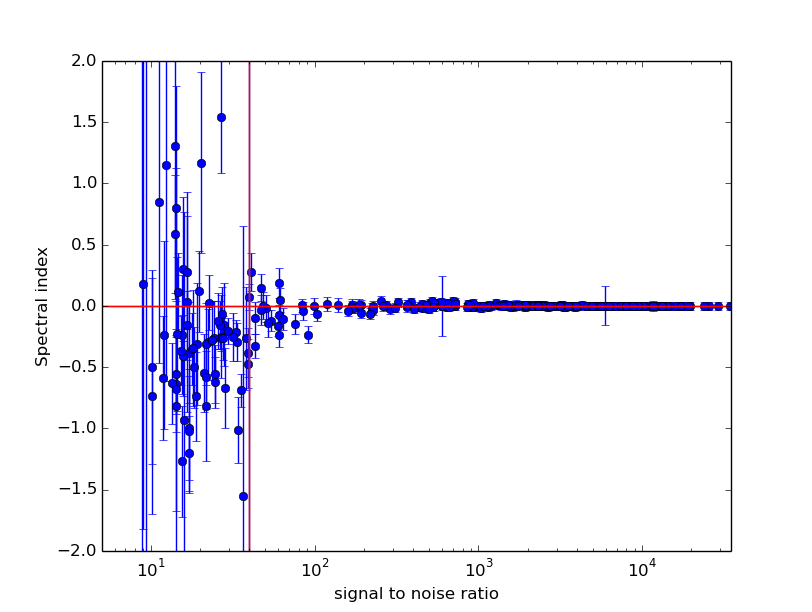}
	\end{center}
	
	\caption{Spectral indices against S/N in the simulation. All the sources have been built to have a flat SED, but the spectral indices are not accurate enough below $40\sigma$ (indicated by the red line).}\label{simulation}
\end{figure}

Since the sub-band maps have a lower S/N than the all-band map, some of the faint sources are detected in less than 2 sub-bands and hence no spectral index can be derived. In the end we extracted the SED for 510 sources of the 614 detected, of which 74 have no rejected frequency points. However not all of the 510 spectral indices are accurate due to the errors on the flux density extraction in the sub-bands. To define the criteria that only select sources with an accurate spectral index, we ran a simulation. We used the \texttt{uvgen} task in \texttt{MIRIAD} \citep{Miriad} to simulate 10 ATCA fields with a total of 301 point sources with $\alpha=0$, and a flux density almost uniformly distributed between 500 $\mu$Jy and 200 mJy at a the same frequency and with the same bandwidth of SCORPIO. We mapped the whole field and derived the spectral index by following the same method used for the SCORPIO data. As shown in Fig. \ref{simulation}, simulated sources with a S/N smaller than $\sim40$ can have a spectral index error greater than 0.5. Therefore in the following analysis we will only refer to the sources brighter than $40\sigma$. 

We also tried to divide the band in 3 sub-bands instead of 7. We found out that the standard deviation of the spectral indices of sources with a S/N greater than 40 and smaller than 100 is 0.2 in the case of 3 sub-bands, 0.03 in the case of 7 sub-bands. This is due to the smaller number of points, leading to a worse linear fit. In the 3 sub-bands case we should choose a 100$\sigma$ threshold. Therefore we did not use the 3 sub-bands.

\begin{figure}
	\begin{center}
		\includegraphics[scale=0.45]{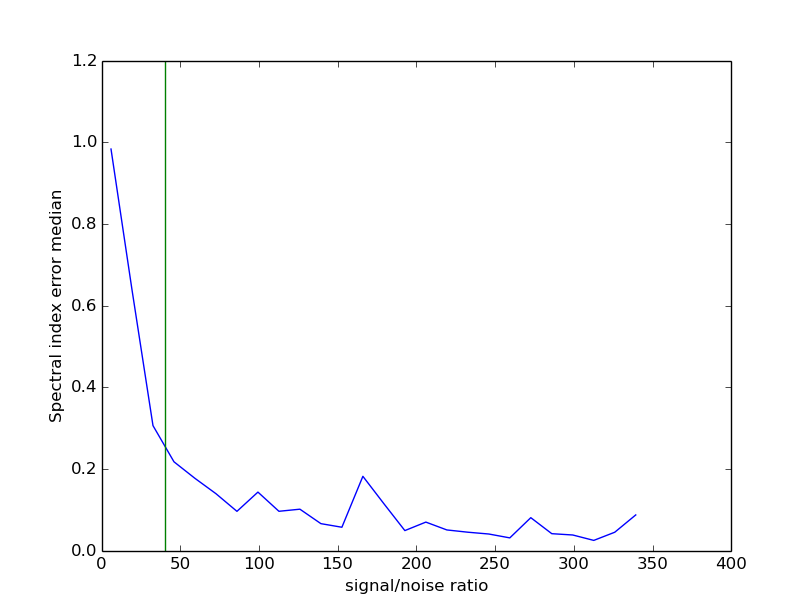}
	\end{center}
	
	\caption{The median on the spectral indices error against S/N in SCORPIO. The green line indicates the $40\sigma$ limit.}\label{erroriindice}
\end{figure}

This limiting S/N is also illustrated by Fig. \ref{erroriindice}. The median of the spectral index errors depends on the S/N as an hyperbola and beyond $\sim40\sigma $ it decreases rapidly. At $40\sigma$ the median on the spectral index errors is $\sim 0.2$, thus it is assumed as the minimal value necessary for a good separation. Our threshold of $40\sigma$ agrees with results by \citet{Rau} that suggest a value between $16\sigma$ and $100\sigma$ as the limit of reliability for spectral indices. Finally, considering that in the pilot field the rms is $\sim 25-30\, \mathrm{\mu}$Jy/beam (Paper I), we adopted a lower limit of 1 mJy for the brightness of the sources. Imposing this limit fixes the number of our sources to 306. 

In the Paper I catalogue we list only 260 sources with flux densities greater than 1 mJy. This discrepancy has two causes. The first is the difference in the flux densities due to the updated primary beam model. The second is the different method used, in the catalogue and in this paper, to extract the flux density. To have a consistent way to extract it in the SCORPIO-ATLAS comparison, we decided to use, as our measured flux densities, the value of the linear fit of the SEDs at 2.1 GHz. As noticed in Sec. \ref{fluxextractor}, the differences in methods and primary beam correction can lead to a slightly higher flux density in our measures compared to the \citet{Franzen} ones.
\subsection{Results}\label{results}
\begin{figure}
    \begin{center}
		\includegraphics[scale=0.45]{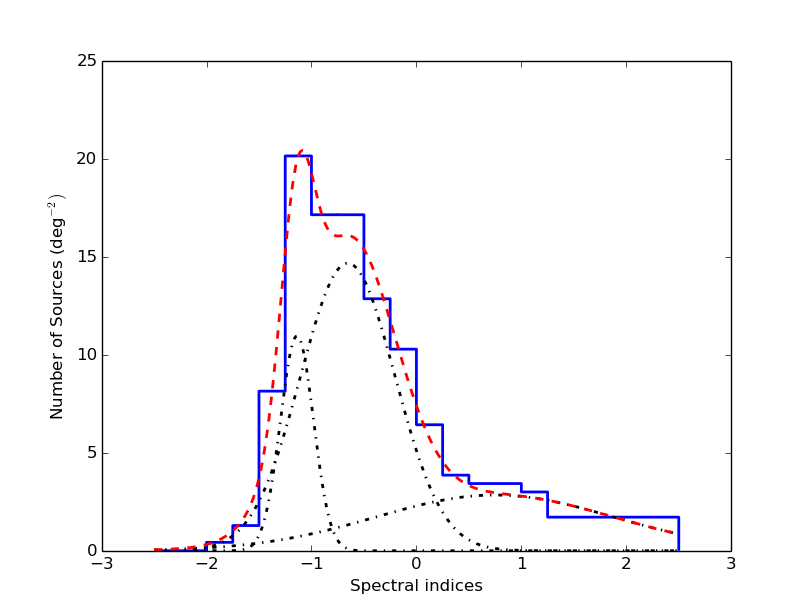}
			\end{center}
			\caption{Spectral index distribution of the SCORPIO survey using a 0.25 wide bin in blue straight line, modeled using 3 Gaussians (red dashed line). In black dash-point line the three Gaussian components of the final model.}\label{SCORPIOisto}
			
		\end{figure}
In Fig. \ref{SCORPIOisto} we show the source density distribution of all the 306 sources brighter than 1 mJy as a function of their spectral indices. The distribution is asymmetric, with a main peak at $\alpha\sim-0.9$ and a long tail at $\alpha\gtrsim0$.

We modelled the spectral index distribution as the sum of several populations, both Galactic and extra-galactic, and fit it with a variable number of Gaussians. We used the Bayesian Information Criterion (BIC) \citep{Schwarz} to select the best-fit model, which will have the smallest BIC. We used the BIC formula as defined in \citet{BIC}:
\begin{eqnarray}
	\mathrm{BIC}=-2\cdot\ln{\hat{L}}+k\cdot\ln{n}
\end{eqnarray}
where $\hat{L}$ is the maximised value of the likelihood function, $n$ is the number of data points and $k$ is the number of free parameters of the model. We used the criteria described in \citet{Kass} to exclude models (see Table \ref{DBIC}). The BIC parameter penalises fits with a larger number of parameters so we can be confident in choosing models with a high number of parameters. Table \ref{BICres} shows that a model consisting of 3 Gaussians (see Fig. \ref{SCORPIOisto}) has a better BIC than the other considered models, such as combinations of any number of Gaussians from 1 to 5 and a skewed Gaussian. The latter is a function similar to an asymmetric Gaussian due to a ``skew'' parameter, defined as $f\left(t\right)=A\phi\left(t\right)\Phi\left(kt\right)$, where $A$ is the amplitude parameter, $k$ is the skew parameter, $\phi\left(t\right)$ is the normal distribution and $\Phi\left(t\right)=\int_{-\infty}^{t}\phi\left(x\right)dx$ is the cumulative distribution function. 

\begin{figure}
    \begin{center}
		\includegraphics[scale=0.45]{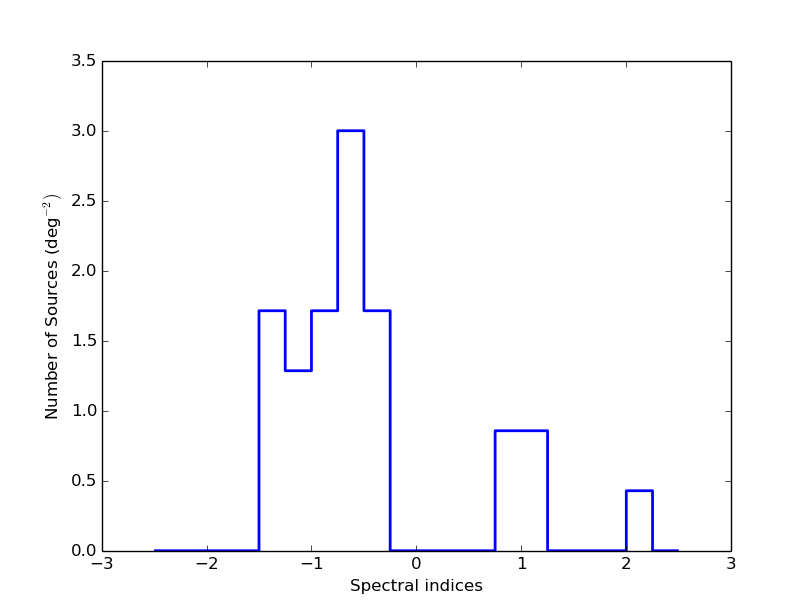}
			\end{center}
			\caption{Spectral index distribution of the SCORPIO group sources using a 0.25 wide bin.}\label{SCORPIOgroup}
			
		\end{figure}

As reported in Sec. \ref{group} we detected 65 group of components that we treated as components of 31 sources. With the same method used for single sources, we checked if they have a different spectral index distribution with respect to the former (Fig. \ref{SCORPIOgroup}). We have a total of 11.5 group of components per square degree with a flux density greater than 1 mJy. Among them, $\sim9.5$ deg$^{-2}$ have a negative spectral index, 1.6 deg$^{-2}$ have a spectral index around 1 and 0.4 deg$^{-2}$ around 2. These numbers are compatible with the all-source results but our sample is too small to make a strong statement.

In the following we compare the statistical results of the spectral index analysis on the SCORPIO data with those of ATLAS, a survey performed at high Galactic latitude. ATLAS can be used as a template for the extra-galactic population of SCORPIO, as the Galaxy is optically thin at radio wavelengths and the cosmological distribution of galaxies is isotropic at these scales.

\begin{table}
	\begin{center}
		\begin{tabular}{l|l} 
		  \hline
			$\Delta\mathrm{BIC}$ & Evidence against higher BIC \\
			\hline
			$0<\Delta\mathrm{BIC}<2$ & Marginal\\
			$2<\Delta\mathrm{BIC}<6$ & Positive\\
			$6<\Delta\mathrm{BIC}<10$ & Strong\\
			$\Delta\mathrm{BIC}>10$ & Very Strong\\
			\hline
		\end{tabular}
		\caption{Criteria to exclude models based on their higher BIC in respect with other models as in \citet{Kass}. }\label{DBIC}
	\end{center}
	\end{table}
\begin{table}
	\begin{center}
		\begin{tabular}{l|l}
			\hline
			Model & BIC\\
			\hline
			Gaussian & 146\\
			Two Gaussians & 96.7\\
			Three Gaussians & 71.7\\
			Four Gaussians & 80.7\\
			Five Gaussians &83.3\\
			One Gaussian plus one skewed Gaussian & 159\\
			Two Gaussians plus one skewed Gaussian &112\\
			\hline
		\end{tabular}
		\caption{BIC for every model considered for the SCORPIO survey. Every Gaussians needs 3 parameters while the skewed Gaussians need 4.}\label{BICres}
	\end{center}

\end{table}
\subsection{Comparison with ATLAS} \label{ATLAS}
\begin{figure}
	\begin{center}
		\includegraphics[scale=0.45]{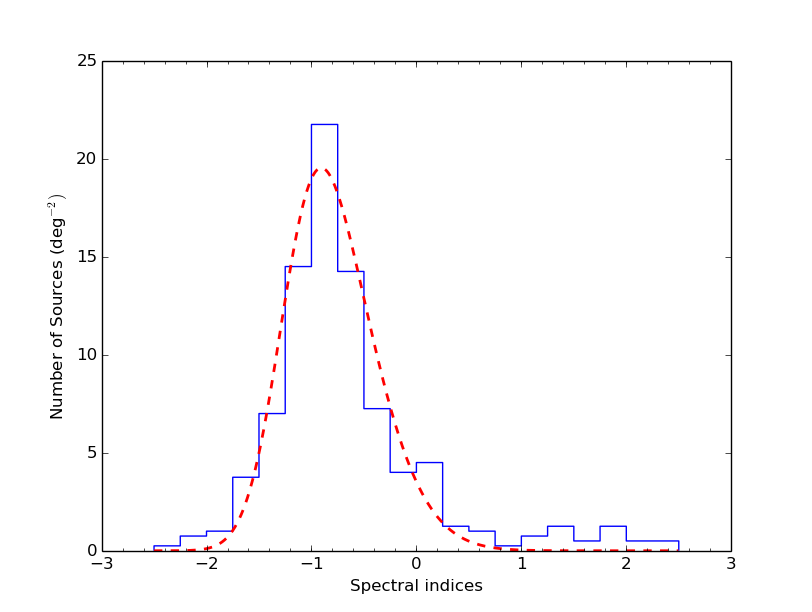}
	\end{center}
	\caption{Number of sources per square degree in a 0.25 wide bin histogram for the ATLAS survey, modelled as a skewed Gaussian (red dashed line).}	\label{ATLASisto}
	
\end{figure}
ATLAS \citep{ATLAS} is a survey conducted on two fields, the Chandra Deep Field-South (CDF-S) and the European Large Area ISO Survey-South 1 (ELAIS-S1) at 1.4 GHz with ATCA. The third ATLAS data release \citep{FranzenAt} includes observations taken using the CABB receiver with a 500 MHz bandwidth covering 1.3-1.8 GHz. Most importantly, the fields are far away from the GP, thus we expect to find only an extra-galactic population.

In \citet{FranzenAt}, the ATLAS DR3 CABB data were divided into two sub-bands and two separate mosaics of each field were created, one using the lower sub-band centred at 1.40 GHz and the other one using the higher sub-band centred at 1.71 GHz. For consistency with SCORPIO, we extracted the spectral indices of the ATLAS sources running the algorithm described in Section \ref{alg} on the two mosaic maps. Then we normalised the number of sources per square degree for both the surveys and only considered the sources brighter than 1 mJy at 2.1 GHz (extrapolated using the spectral index). As shown in Fig. \ref{ATLASisto}, the synchrotron peak is the predominant component in ATLAS. There is a minor peak around 0, which can also be explained by extra-galactic sources, the Gigahertz Peaked Spectrum (GPS) sources (GPS; \citealt{Odea}), that can have the turnover frequency between 1.4 and 1.71 GHz. We tried to fit this distribution with different models, as in Section \ref{results}, and we found that the Gaussian model has a BIC value of 85.4 , higher than the skewed gaussian one (81.7). Consequently the latter is the one we selected (see Fig. \ref{ATLASisto} for the model). Assuming that virtually all the ATLAS sources are extra-galactic, we used the ATLAS model as a template for the extra-galactic population of SCORPIO. 

\begin{figure}
	\begin{center}
		\includegraphics[scale=0.45]{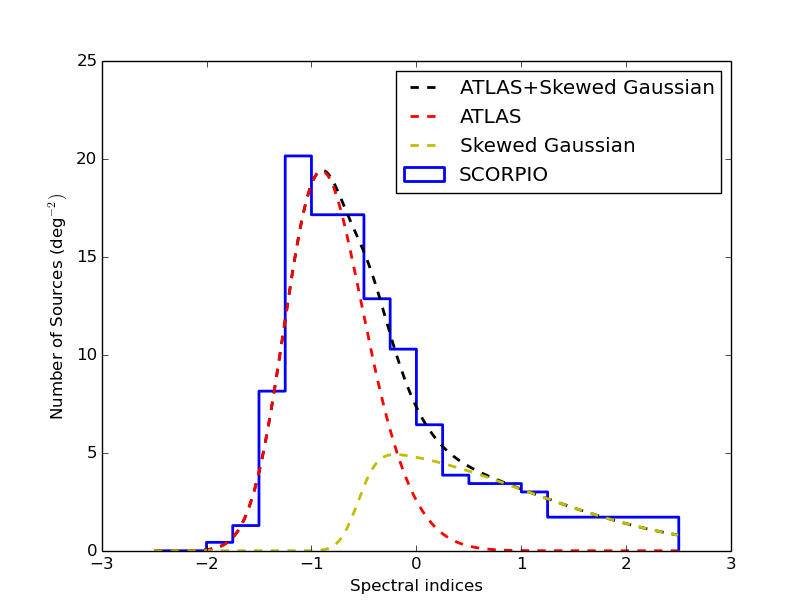}
	\end{center}
	\caption{SCORPIO spectral index distribution modelled as a sum of the ATLAS model and a skewed Gaussian.}	\label{SCORPIOmodel}
\end{figure}

We found 20 percent more sources per square degree in SCORPIO than in ATLAS (120 versus 100) and we found that a model using the ATLAS skewed Gaussian plus another skewed Gaussian gives a BIC of 76.8, better than almost all the models shown in Table \ref{BICres} (with the exception of the one that uses three Gaussians). Nevertheless the model including the ATLAS extra-galactic template has, as anticipated before, a physical motivation making it the more reliable one. Fig. \ref{SCORPIOmodel} shows that the ATLAS Gaussian fits quite well the peak at $\alpha=-0.9$ and, most importantly, another skewed Gaussian fits the other peaks. Maybe larger statistics could help us in concluding that they are part of more than one galactic population, as expected, but, with our data, we can only say that they are part of a generic galactic population.

\begin{figure}
	\begin{center}
		\includegraphics[scale=0.45]{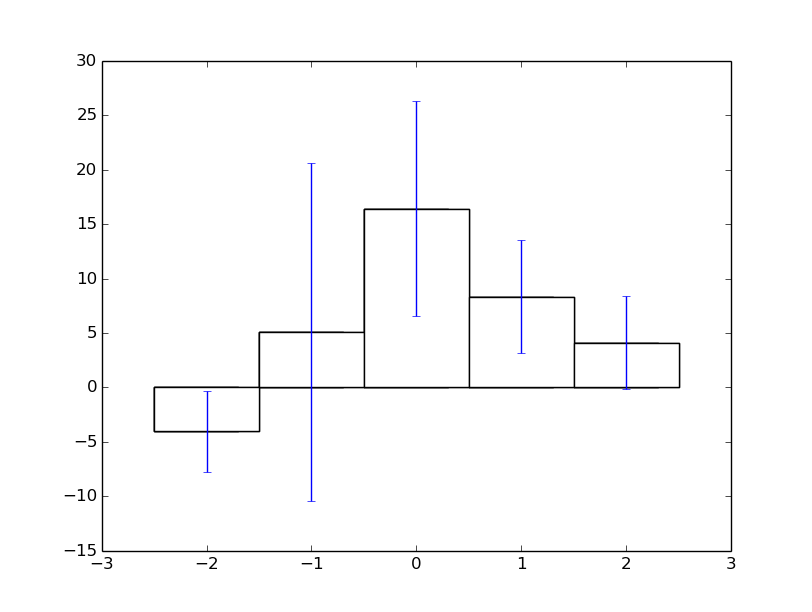}
	\end{center}
	\caption{The difference between the SCORPIO and the ATLAS spectral indices histogram.}	\label{sottrazione}
\end{figure}

In Fig. \ref{sottrazione} the difference between the SCORPIO and the ATLAS population per spectral index bin is shown. The error $e_\mathrm{i}$ associated with the $i^{th}$ bin is:
\begin{eqnarray}
	e_\mathrm{i}=\sqrt{N_\mathrm{Si}}+\sqrt{N_\mathrm{Ai}}
\end{eqnarray}
where $N_\mathrm{Si}$ and $N_\mathrm{Ai}$ are respectively the number of SCORPIO sources and ATLAS sources in the $i^{th}$ bin. We can recognize 3 regions of the spectral index $\alpha$:
\begin{enumerate}
	\item $-2.5<\alpha<-0.5$: the number of excess sources is not significantly different from zero, implying that all SCORPIO sources in this spectral index range may be extra-galactic. In this spectral index range there can also be unresolved or almost unresolved Young Supernovae Remnants (YSNRs). To know what size we should expect from different age SNRs, we consider the size of the SN1987A remnant. The SN1987A remnant is located in the Large Magellanic Cloud, at $\sim51.4$ kpc from the Sun. It expanded from 0.21 to 0.39 pc between 1995 and 2010 with an almost linear expansion rate of about 0.01 pc/year \citep{Chiad}. Therefore a 100 years old similar supernovae would have a linear dimension of $\sim1.1$ pc and an angular dimension of $\sim11''$ at a 20 kpc distance from the Sun. Given the $14.0''\times6.5''$ resolution of SCORPIO, we may not be able to resolve it.
	
We now estimate how many YSNRs are expected. We assume that in the last 1000 years there were 7 supernovae in our Galaxy within 5 kpc from the Earth in the GP. Given the position of the Sun in the GP, from the point of view of the Galactic Centre this corresponds to a $\sim$50 deg circular sector. Assuming that the distribution is uniform in the GP, we estimate about one supernova every $\sim$20 years in our Galaxy. Considering that $\sim66$ percent of the OB stars lies in the direction of the Galactic Center \citep{Robin} in a 120 deg$^2$ area, we can assume a supernova every $\sim$40 square degree younger than 100 years in the direction of the SCORPIO field.

Moreover we have to consider that SN discovery is tipically performed at optical wavelength. Thus, even if the remnants of these supernovae are bright in radio (at peak we expect almost 1000 Jy in $L$-band at 20 kpc \citealt{bufano}), they could be very faint in the optical due to the GP absorption: considering that $m = A + M + \mu$, assuming a SN with a peak absolute magnitude of $M = −17$ mag, we would have an apparent magnitude as high as $m\sim$26 mag, having in the direction of SCORPIO an extinction A = 26.7 \citep{Schlafly} and a distance module, corresponding to 20 kpc, of $\mu\sim16.5$. This means that it is possible for a supernova in our Galaxy to happen undiscovered but to leave a remnant bright enough to be detected at radio wavelength. Even with this premise, considering a density of 0.025 supernova younger than 100 years per square degree in the direction of SCORPIO, we would not be able to detect more than one YSNR in the relatively small SCORPIO field of view ($\sim5$ square degree for the whole field), confirming the extragalactic origin of the sources in this $\alpha$ range;
	\item $-0.5<\alpha<0.5$: the excess here is $\sim16\pm10$ per square degree. These are typical spectral indices of Galactic source that show thermal emission in an optically thin environment, e.g. H\textsc{ii} regions, Planetary Nebulae (PNe) and Luminous Blue Variable (LBV) (e.g. \citealt{Paladini}, \citealt{Cerrigone}, \citealt{Agliozzo12}, 2014, \citealt{Ingallinera}). 
	\item $0.5<\alpha<2.5$: there is a significant excess of about $12\pm8$ sources per square degree with a spectral index in this range, suggesting stellar winds, interacting stellar winds and thermal emission in an optically thick environment, e.g. Wolf Rayet, LBV, OB stars and compact and ultracompact H\textsc{ii} regions (e.g. \citealt{Scuderi}, \citealt{Umana}, \citealt{Leto}).
\end{enumerate} 
Obviously we expect some overlap between spectral index regions. Furthermore, the different frequency range of the ATLAS survey makes it likely to have a surplus of extra-galactic flat sources due to possible turnover positions. 

Note that the difference between the total sources surplus in SCORPIO (20 percent) and the one in the interval $-2.5<\alpha<2.5$ is due to the smaller bandwidth in ATLAS, resulting in a larger number of spurious spectral indices.

We cross-matched our point sources with the \citet{Anderson} H \textsc{ii} regions catalogue, extracted from infrared colours, and we found 7 matching sources, reported in Table \ref{anderhii}: 4 of them do not have a counterpart in the Southern Galactic Plane Survey (SGPS, \citealt{Mcclure}), suggesting that many of the ``radio quiet'' H \textsc{ii} regions emit in radio being actually ``radio weak''; the remaining 3 are H \textsc{ii} region candidates. The spectral indices of SCORPIO1\_123b, SCORPIO1\_169, SCORPIO1\_218 and SCORPIO1\_483 are consistent with thermal free-free emission in an optically thick medium in the first case, and in an optically thin medium in the latter ones. The other 3 sources present a much less reliable spectral index because they are resolved and embedded in diffuse emission.
\begin{table}
	\begin{center}
		\begin{tabular}{lrrl}
			\hline
			SCORPIO  &Flux Density& Spectral& Catalogue\\
			ID&(mJy)& Index&Tag\\
			\hline
			SCORPIO1\_123b&$40.5\pm1.3$&$1.1\pm0.1$&Q\\
			SCORPIO1\_136&$0.7\pm0.1$&$-0.7\pm0.6$&Q\\
			SCORPIO1\_169&$24.5\pm1.3$&$-0.34\pm0.13$&Q\\
			SCORPIO1\_218&$182\pm5.5$&$-0.09\pm0.05$&C\\
			SCORPIO1\_219&$9.2\pm0.7$&$-1.0\pm0.4$&C\\
			SCORPIO1\_459&$4.7\pm0.2$&$-1.5\pm0.4$&Q\\
			SCORPIO1\_483&$21.5\pm0.7$&$-0.12\pm0.04$&C\\	
			\hline
		\end{tabular}
		\caption{SCORPIO ID, flux density, spectral index and catalogue tag on \citet{Anderson} of the 7 crossmatched sources. The catalogue tags stand for radio Quiet and for Candidate, indicating respectively a H \textsc{ii} region detected in the infrared but without a counterpart in the radio catalogue considered in \citet{Anderson}, the SGPS and a bubble candidate to be a H \textsc{ii} region.}\label{anderhii}
	\end{center}

\end{table}

 \section{Summary and conclusions}
We have described:
\begin{enumerate}
	\item An algorithm, based on the CASA task \texttt{imfit}, to automatically extract the flux density of a large number of point-like sources by fitting Gaussians and to calculate their spectral indices;
	\item The application of this algorithm to the SCORPIO data to derive the SCORPIO sources spectral index distribution, along with the discussion about the different source populations that contribute to the distribution;
	\item The comparison between SCORPIO and an extra-galactic survey, ATLAS. 
\end{enumerate}
We have found that, with a bandpass like the SCORPIO one, the S/N for the object in the field has to be at least 40 in order to rely on the spectral index $\alpha$ with an uncertainty $\lesssim0.2$. 

We have found that, in the Galactic Plane, the source count with flux densities greater than 1 mJy is about 20 percent higher than the source count at high galactic latitude. We found 16 Galactic sources per square degree with a spectral index of about $\alpha=0$, suggesting optically thin thermal emission such as H\textsc{ii} regions and planetary nebulae, while the remaining 12 sources per square degree present a spectral index $0.5<\alpha<2.5$, pointing to an optically thick thermal emission such as stars and compact H\textsc{ii} regions.

These results are very important for planning forthcoming radio surveys. This work will be eventually extended to the whole SCORPIO field. Considering that the survey will be $\sim5\,\mathrm{deg}^2$ and Galactic source density above 1 mJy of 20 deg$^{-2}$ as reported in this paper, we expect about 100 Galactic sources brighter than 1 mJy. Thanks to the larger sample that we will consider, we will be able to confirm or not the results of this paper, validating their applications to the future surveys.

\appendix
\section{Catalogue}
We created a new catalogue of the spectral indices of the 510 sources listed in paper I for which we could calculate it (see Table \ref{catal} for the first 10 entries). The columns are:
\begin{enumerate}
\item ID of the source in the SCORPIO catalogue;
\item galactic longitude;
\item galactic latitude;
\item flux density at 2.1 GHz;
\item error on the flux density (see details on how we calculated that on Paper I)
\item spectral index;
\item error on the spectral index.
\end{enumerate}
\begin{table*}
	\begin{center}
		\begin{tabular}{l r r r r r r} 
		\hline
		SCORPIO ID& l~~~~~~~&b~~~~~~& S~~~~& $\Delta S~~$ &$\alpha~~$& $\Delta\alpha~~$\\
		& (deg)~~~ &(deg)~~~& (mJy)& (mJy)\\
		  \hline
SCORPIO\_001 & 343.0025 &1.7604& 33.33 &1.03 &-0.39& 0.07 \\
SCORPIO\_002 & 343.0051 &0.2234& 0.49 &0.09 &6.50 &3.14 \\
SCORPIO\_003 & 343.0134 &0.6086& 7.74 &0.29 &-0.74 &0.18 \\
SCORPIO\_004 & 343.0138 &1.7208 &1.60 &0.17 &-2.80& 1.45 \\
SCORPIO\_006 & 343.0152 &0.1166 &1.27 &0.16 &0.02 &0.85 \\
SCORPIO\_007 & 343.0157 &-0.1830& 1.51 &0.12& -0.32 &0.39 \\
SCORPIO\_005b& 343.0186 &1.1577 &106.56& 3.20& -0.40& 0.10 \\
SCORPIO\_010 & 343.0203 &0.3109 &20.97& 0.64 &-1.04& 0.03 \\
SCORPIO\_011 & 343.0216 &0.7407 &1.12 &0.10 &-1.50 &0.73 \\
SCORPIO\_012 & 343.0277& 0.8702 &1.17& 0.09 &0.65 &0.35 \\
\hline
		\end{tabular}
		\caption{First 10 entries of the spectral indices catalogue.}\label{catal}
	\end{center}
	\end{table*}

\end{document}